\documentclass[
  a4paper, 
  12pt, onecolumn,
]{article}

\title{\bf Spectral Degeneracies in the Totally Asymmetric Exclusion Process
}
\author{      O. Golinelli, K. Mallick
\bigskip
\\ \ad        Service de Physique Th\'eorique, Cea Saclay, 91191 Gif, France
}
\date{\normalsize       
                        December 16, 2004
\\                      Preprint T04/167; arXiv:cond-mat/0412462
}

\newcommand  {\ad}{\normalsize\em}      
\usepackage[final]{graphicx}
\newcommand{\figwidth}{\columnwidth}
\newcommand{\packagewidth}{1.5truecm}

\begin{document}
\maketitle

\begin{abstract}
  \normalsize

We study the spectrum of the Markov matrix of the totally asymmetric
exclusion process (TASEP) on a one-dimensional periodic lattice at {\it
arbitrary} filling.  Although the system does not possess obvious
symmetries except translation invariance, the spectrum presents many
multiplets with degeneracies of high order.  This behaviour is explained
by a hidden symmetry property of the Bethe Ansatz.  Combinatorial
formulae for the orders of degeneracy and the corresponding number of
multiplets are derived and compared with numerical results obtained from
exact diagonalisation of small size systems.  This unexpected structure
of the TASEP spectrum suggests the existence of an underlying large
invariance group.


\medskip \noindent 
Pacs number: 05.40.-a, 05.60.-k

\end{abstract}

\section{Introduction}

The asymmetric simple exclusion process (ASEP) is a driven lattice gas
model in which particles interact by hard core exclusion. This simple
system has been introduced as a building block for models of hopping
conductivity, motion of RNA templates, traffic flow and surface growth.
  From a theoretical point of view,
the ASEP plays a fundamental role in the study of non-equilibrium
processes: many exact results have been obtained concerning
one-dimensional phase transitions (Derrida et al. 1993), phase
segregation (Evans et al. 1998), large deviations functions and
fluctuations far from equilibrium (Derrida et al. 2003). For  a review,
see Derrida (1998),  Sch\"utz (2001).  In the absence of a driving field,
the {\it symmetric} exclusion process can be mapped into the
Heisenberg spin chain.  The asymmetry due to a non-zero external
driving field breaks the left/right symmetry and the ASEP is
equivalent to a non-Hermitian spin chain of the XXZ type. The ASEP can
also be mapped into a two-dimensional six-vertex model at equilibrium.
These mappings allow to use the methods of integrable
systems, such as the Bethe Ansatz (Dhar 1987, Gwa and Spohn 1992, Kim
1995).

In a recent article (Golinelli and Mallick 2004), we carried out a
spectral study of the Markov matrix of the exclusion process on a
periodic lattice. At half-filling the system is invariant under charge
conjugation combined with reflection in addition to being translation
invariant.  We showed that these symmetries predict the existence of
singlets and doublets in the spectrum. However, for the {\it totally}
asymmetric simple exclusion process (TASEP) the spectral structure is
much richer.  We observed numerically that unexpected degeneracies of
higher order exist and become generic as the system size increases. We
explained, in an heuristic manner, the existence of these degeneracies
by using the fact that some of the solutions of the Bethe equations
appear in pairs with opposite values.

In the present work, we generalize our previous study to the TASEP at
{\it arbitrary filling}.  We perform an exhaustive analysis of the
spectral degeneracies of the TASEP on a periodic lattice. The orders
of degeneracies observed and the number of multiplets with a given
degeneracy depend on commensurability relations between the number of
sites and the number of particles. Although the Bethe equations do not
exhibit any obvious symmetry, we find that they possess an invariance
under exchange of roots that allows to predict combinatorial formulae
for the orders of degeneracies and the number of multiplets of a given
order of degeneracy.  Our formulae are confirmed by direct numerical
diagonalisation of small size systems. This peculiar structure of
the TASEP spectrum suggests the existence of some underlying
symmetries of the model that may shed light on its remarkable
combinatorial properties.

The outline of this article is as follows. In section 2, the
definition and basic properties of the TASEP are recalled. In section
3, we present a self-contained derivation of the Bethe equations based
upon the fact that the Bethe wave function is a determinant for the
TASEP model.  The symmetries of the Bethe equations are studied in
section 4 and combinatorial expressions for the degeneracies  are
derived in section 5. In section 6, we discuss the behaviour of large
size systems and present numerical results.  Concluding remarks appear
in section 7.  In the Appendix, the geometrical setting of the roots
in the complex plane is briefly described.

\section{The TASEP model}

The simple exclusion process is a stochastic process in which
particles hop on a lattice and respect the {\em exclusion rule} that
forbids two or more particles per site.  On a one-dimensional lattice,
this rule prohibits overtaking between particles.

In this article we consider a periodic 1-d lattice of length $L$,
i.e., a ring where sites $i$ and $i+L$ are identical.  The system is
closed and the number $N$ of particles is fixed with $N \le L$.  The
{\em filling} (or density) is given by $\rho = N/L$.

The particles evolve with the following dynamics rule: during the time
interval $[t, t+dt]$, a particle on a site $i$ jumps with probability
$dt$ to the neighbouring site $i+1$ if it is vacant.  As the jumps are
allowed in only one direction, the model considered is the {\em
  totally asymmetric} exclusion process (TASEP).

A configuration ${\cal C}$ of the system is characterised by the list
of the $N$ occupied sites amongst the $L$ available sites.  The total
number of configurations is therefore
\begin{equation}
  \Omega = {L \choose N} = \frac{L!}{N! (L-N)!}   \, .
 \label{eq:defomega}
\end{equation}

Let $\psi_t({\cal C})$ be the probability that the configuration of
the system at time $t$ is ${\cal C}$.  As the TASEP is a
continuous-time Markov (i.e., memoryless) process, the
$\Omega$-dimensional vector $\psi_t$ evolves according to the {\em
  master equation}
\begin{equation}
  \frac{d\psi_t}{dt} = M \psi_t  \, , 
\end{equation}
where $M$ is the $\Omega \times \Omega$ Markov matrix.  For ${\cal C}
\ne {\cal C'}$, the term $M({\cal C'},{\cal C})$ represents the
transition rate from ${\cal C}$ to ${\cal C'}$: it is equal to 1 if
${\cal C'}$ is obtained from ${\cal C}$ by an allowed jump of one
particle, and is 0 otherwise. The element $-M({\cal C}, {\cal C})$ is
equal to the number of allowed jumps from ${\cal C}$. Thus,  the sums 
columns  of the Markov matrix vanish and the total probability is conserved. 
 For the exclusion process  on a periodic
lattice, the sums over lines of $M$ also vanish: the stationary
probability (which corresponds to the eigenvalue 0) is thus uniform:
$\psi({\cal C}) = 1/\Omega$.

As the dynamics is  ergodic and aperiodic, $M$ satisfies the
conditions of the Perron-Frobenius theorem: the eigenvalue 0 is
non-degenerate and all the other eigenvalues have strictly negative
real parts (equal to the inverse of the relaxation times).  As $M$ is
a real non-symmetric matrix, the eigenvalues are either real numbers
or complex conjugate pairs.
 
In this work, we shall investigate the spectral degeneracies, i.e.,
equalities amongst the eigenvalues of the Markov matrix.

\section{Derivation of the Bethe equations}

Since the work of Dhar (1987), it is known that the {\em  Bethe  Ansatz}
can be applied to the ASEP.  In this section, we give a self-contained
derivation of the Bethe equations for the particular case of the
TASEP, much simpler than that of the generic ASEP (Gwa and Spohn
1992).

A configuration ${\cal C}$ will be represented by the sequence $(x_1,
x_2,\dots, x_N)$, the integers $x_i$ being the positions of the
particles with
\begin{equation}
    1 \le x_1 < x_2 < \dots < x_N \le L   \, .
    \label{eq:cond}
\end{equation}
The idea of the Bethe {\em Ansatz} consists
  in writing the eigenvectors   $\psi$  of
the Markov matrix as linear combinations of plane waves
(see, e.g.,  Gaudin 1983).  In fact, the
Bethe wave function $\psi$ of the TASEP is a determinant (Gaudin and
Pasquier 2004).  We therefore define $\psi$ as
\begin{equation}
    \psi(x_1,\dots,x_N) = \det(R)  \, , 
   \label{eq:psidet}
\end{equation}
where $R$ is a $N \times N$ matrix with elements
\begin{equation}
   R(i,j) = \frac{z_{i}^{x_j}}{(1-z_i)^j} \ \ \mbox{for } 1 \le i,j \le N  \, ,
   \label{eq:r}
\end{equation}
$(z_1, \dots, z_N)$ being $N$ given complex numbers.  If we assume
that $\psi$ is of this form and that it is an eigenvector of $M$, the
$z_i$'s then must satisfy some conditions, the {\em Bethe equations},
 that we now re-derive. 

We first show that $\psi$ defined by equations~(\ref{eq:psidet},
\ref{eq:r}) satisfies two identities which are valid for any values of
$z_i$ and of $x_i$, even without imposing the ordering given in
equation~(\ref{eq:cond}).  The first identity is
\begin{equation}
  E \psi(x_1,\dots, x_N) = \sum_{k=1}^N
   [\psi(x_1,\dots,x_k-1,\dots, x_N) - \psi(x_1,\dots,x_k,\dots, x_N)] \, , 
  \label{eq:evp}
\end{equation}
for any $(x_1, \dots, x_N)$ and $(z_1, \dots, z_N)$, and where $E$ is
given by
\begin{equation}
   E = -N + \sum_{i=1}^N 1/z_i 
   \, .
   \label{eq:e}
\end{equation}
Equation~(\ref{eq:evp}) is obtained by writing
\begin{eqnarray}
  \psi(x_1,\dots,x_k-1,\dots, x_N) - \psi(x_1,\dots,x_k,\dots, x_N) 
   =  \nonumber \\
  \det \left( R(i,1) , \dots ,\left( \frac{1}{z_i} -1 \right) R(i,k) ,
           \dots , R(i,N)
        \ \right) \, .
\end{eqnarray}
This determinant is similar to $ \det(R)$ except for the $k$-th
column.  Expanding this determinant over all permutations of $\{ 1,
\dots, N \}$ and performing the sum over $k=1,\dots,N$ leads to
equations~(\ref{eq:evp}, \ref{eq:e}).

The second identity valid for any $(z_1, \dots, z_N)$ and any $(x_1,
\dots, x_N)$ with $x_{k-1} = x_k$ (two  particles collision), is
\begin{equation}
  \psi(x_1,\dots, x_k, x_k, \dots, x_n) - 
  \psi(x_1,\dots, x_k, x_k+1, \dots, x_n)  = 0 \, .  
  \label{eq:kk+1}
\end{equation}
The left hand side of equation~(\ref{eq:kk+1}) can be written as
$\det(\tilde{R})$ where $\tilde{R}$ is a matrix that is identical to
$R$ but for its $k$-th column that is given by
\begin{equation}
  \tilde{R}(i,k) = \frac{z_i^{x_k} - z_i^{x_k+1}}{(1-z_i)^{k}} = 
                \frac{z_i^{x_k}}{(1-z_i)^{k-1}} = R(i,k-1) =  \tilde{R}(i,k-1)
\end{equation}
The $(k-1)$-th and the $k$-th columns of $\tilde{R}$ are equal and,
therefore, $\det(\tilde{R}) = 0$. This proves equation~(\ref{eq:kk+1}).

The eigenvalue equation, $E\psi = M \psi$, is written as
equation~(\ref{eq:evp}) except that the sum is restricted to the
allowed jumps of particles, i.e., to the indices $k$ such that
$x_{k-1} +1 < x_{k}$.  However, in equation~(\ref{eq:evp}), the terms
with $x_{k-1}+ 1 = x_{k}$ vanish thanks to equation~(\ref{eq:kk+1}).
Thus equation~(\ref{eq:evp}) is identical to the eigenvalue equation
if the eigenvector has the form assumed in equations~(\ref{eq:psidet},
\ref{eq:r}).

The function $\psi$ must also satisfy periodic boundary conditions
\begin{equation}
  \psi(x_1, x_2, \dots, x_n) = \psi(x_2, \dots, x_n, x_1 + L) \, .
  \label{eq:cycl}
\end{equation}
The  periodic conditions are the ones that
 quantify the eigenvalues by imposing  a set of equations on the $z_i$'s,
 the   Bethe equations.  Denoting by $i$ and $j$ the generic line
and column indices of the matrices, respectively, we have
\begin{equation}
  \psi(x_2, \dots, x_N, x_1 + L) = \det \left( 
       \frac{z_{i}^{x_{2}}}{1-z_i}, 
     \dots,  \frac{z_{i}^{x_{j+1}}}{(1-z_i)^j} , \dots,
             \frac{z_i^{x_1+L}}{(1-z_i)^N}  \right) \, .
  \label{eq:pbc}
\end{equation}
By cyclic permutation of the columns, we obtain
\begin{eqnarray}  
  & &  \psi(x_2, \dots, x_N, x_1 + L)     \nonumber  \\
    &=&   (-1)^{N-1} 
    \det \left( \frac{z_i^{x_1+L}}{(1-z_i)^N},  \frac{z_{i}^{x_{2}}}{1-z_i}, 
     \dots,  \frac{z_{i}^{x_j}}{(1-z_i)^{j-1}} , \dots   \right)  \nonumber \\
     &=&   (-1)^{N-1} 
    \det \left( \frac{z_i^L}{(1-z_i)^{N-1}} \, R(i,1),
     \dots,  (1-z_i)\, R(i,j) , \dots \right)    \nonumber  \\
    &=&   (-1)^{N-1}  \prod_{k=1}^N (1-z_k)  \,\,\, \
    \det \left( \frac{z_i^L}{(1-z_i)^N} \, R(i,1),
     \dots,  R(i,j) , \dots   \right)     \, .  
  \end{eqnarray}
  This last term is equal to $\psi(x_1, x_2, \dots, x_n) = \det(R) \, $ if
  $z_1,\dots, z_N$ are solutions of the $N$   Bethe equations
\begin{equation}
  (z_i-1)^N z_i^{-L} = - \prod_{k=1}^N(1-z_k) \ \ \
   \hbox{for}    \ \ \  i=1,\dots,N  \, .
 \label{eq:be}
\end{equation} 
The vector $\psi$ defined by equations~(\ref{eq:psidet}, \ref{eq:r})
is then an eigenvector of the Markov matrix $M$ with eigenvalue $E$
given by equation~(\ref{eq:e}).

The uniform stationary probability with $E=0$ corresponds to the very
special solution where all the $z_i=1$.  For all the other solutions
of the Bethe equations, the $z_i$'s are distinct and are different
from 1; hence the determinant $\det(R)$ does not vanish.

\section{Symmetries of  the Bethe equations}

In this section we show that the Bethe equations~(\ref{eq:be}) have
certain solutions that are distinct but lead to the same eigenvalue
$E$.

  Following Gwa and Spohn (1992), we introduce new variables
$\tilde{z_i} = 2/z_i-1$ in the Bethe equations~(\ref{eq:be}) which
then become
\begin{equation}
  (1-\tilde{z_i})^N \ (1+\tilde{z_i})^{L-N} 
  = - 2^L \prod_{k=1}^N \frac{\tilde{z_k}-1}{\tilde{z_k}+1}
  \ \ \ \hbox{for}    \ \ \  i=1,\dots,N  \, .
  \label{eq:bep}
\end{equation} 
The corresponding eigenvalue $E$
is given by
\begin{equation}
   2E = - N  + \sum_{k=1}^N \tilde{z_k}    \, .
   \label{eq:ez}
\end{equation}
We remark that the right hand side of equation~(\ref{eq:bep}) is
independent of $i$. We shall analyse equation~(\ref{eq:bep}) in three
steps.  Firstly, we consider the polynomial equation of degree $L$ for
a given complex parameter $Y$,
\begin{equation}
   (1-Z)^N \ (1+Z)^{L-N} = Y  \, . 
   \label{eq:poly}
\end{equation}
We call $(Z_1, \dots, Z_L)$ the roots of this polynomial.  In the
Appendix, we explain how the $Z_i$'s can be labelled so that for a
given $i$, $Z_i$ is an analytic function of $Y$ in the complex plane
with a branch cut along the real semi-axis $[0, +\infty)$.  Secondly,
we choose a set $c = \{c_1, \dots, c_N \}$ of $N$ distinct indices
among $\{1, \dots, L \}$. The number of possible sets $c$ is precisely
$\Omega$, the total number of configurations
(equation~(\ref{eq:defomega})).  Finally, for a given choice set $c$,
we define a function of $Y$
\begin{equation}
   A_c(Y) = - 2^L \prod_{k=1}^N \frac{Z_{c_k}-1}{Z_{c_k}+1}
   \, .
\end{equation}
The Bethe equations~(\ref{eq:bep}) are now equivalent to the
self-consistency equation
\begin{equation}
   A_c(Y) = Y
   \, .
   \label{eq:acy=y}
\end{equation}
  For a given choice set $c$ and a root $Y_c$ of the last equation, the
$Z_k$'s are determined by equation~(\ref{eq:poly}).  The solutions of
the Bethe equations are then given by $\tilde{z_k} = Z_{c_k}$.  The
corresponding eigenvalue $E_c$ is obtained from equation~(\ref{eq:ez})
\begin{equation}
      2 E_c  = - N   + \sum_{k=1}^N  Z_{c_k}   \, .
 \label{eq:defEc}
\end{equation}
The eigenvector $\psi$ is given by 
  equations~(\ref{eq:psidet}, \ref{eq:r}),  
using $z_i = 2/(\tilde{z_i}+1)$.

In order to understand the origin of the spectral degeneracies, we
must consider the case where $L$ (the number of sites) and $N$ (the
number of particles) are not relatively prime.  We define the integers
$p$, $n$ and $l$ as follows
\begin{equation}
   p = \gcd(L,N) \ , \ \ L = p l\ , \ \ N = p n \ . \label{eq:defp}
\end{equation}
Let $(y_1, \dots, y_p)$ be the $p$-th roots of $Y$ (i.e., $y_k^p = Y$)
labelled as $ 0 \le \arg(y_1) < \dots < \arg(y_p) < 2\pi$.
Equation~(\ref{eq:poly}) is thus equivalent to the $p$ polynomial
equations of degree $l$
\begin{equation}
   Q_k(Z) = (1-Z)^n \ (1+Z)^{l-n} - y_k  = 0 
    \ \ \mbox{for } \  \  k = 1,  \ldots ,   p  \, .
   \label{eq:polyk}
\end{equation} 
Thus the set $\{ Z_1, \dots, Z_L \}$ of the $L$ solutions of
equation~(\ref{eq:poly}) is made up of $p$ {\em packages}, the $k$-th
package being constituted by the $l$ solutions of $Q_k(Z) = 0$.  Let
us call ${\cal P}_k$ the set of indices of the $k$-th package: in
other words the solutions of $Q_k(Z) = 0$ are the $Z_i$ with $i \in
{\cal P}_k$.  The labelling of the $Z_i$ described in the Appendix
shows explicitly that
\begin{equation}
  {\cal P}_k = \{ k, k+p, k+2p, \dots, k+L-p \}  \, , \label{eq:defpack}
\end{equation}
i.e., ${\cal P}_k$ contains the indices $i$ such that $i = k$ modulo
$p$.
\label{s:package}

  For any given package ${\cal P}_k$, we have the fundamental equations
\begin{eqnarray}
  \sum_{i \in {\cal P}_k} Z_i  &=& 2n-l \, , 
  \label{eq:sz}
  \\
  \prod_{i \in {\cal P}_k} \frac{Z_i-1}{Z_i+1} &=& 1  \, .
  \label{eq:pz}
\end{eqnarray}
We emphasize that the right hand sides are independent of $k$.
These identities are derived as follows.  We have $(-1)^n Q_k(Z) =
\prod_{i \in {\cal P}_k} (Z - Z_i)$ because 
 the $ Z_i$'s    with $i \in {\cal P}_k $   are the roots
of the polynomial $Q_k(Z)$.  The evaluation of the coefficient of
$Z^{l-1}$ leads to the equation~(\ref{eq:sz}) except when $l=1$.
Similarly, the evaluation of $Q_k(1)/Q_k(-1)$ yields
equation~(\ref{eq:pz}) except when $n=0$ or $n=l$.  These exceptions
correspond to a trivial model which is either empty, $N=0$, or full,
$N=L$, (the spectrum of $M$ is then reduced to the single eigenvalue
$\{ 0 \}$).  In the following, we assume that $0 < N < L$ and so
equations~(\ref{eq:sz}, \ref{eq:pz}) are true.

We now consider a solution $Y_c$ of the Bethe
equation~(\ref{eq:acy=y}) associated with a given choice set $c$.
Assume that there exists a package ${\cal P}_{f}$ such that $c$
contains ${\cal P}_{f}$ (i.e., ${\cal P}_{f} \subset c$) and that
there exists another package ${\cal P}_{e}$ such that $c$ is disjoint
from ${\cal P}_{e}$ (i.e., ${\cal P}_{e} \cap c = \emptyset$).  We
define a new choice set $c'$ of $N$ indices by exchanging the package
${\cal P}_{f}$ with ${\cal P}_{e}$, i.e., 
 $c'= (c / {\cal P}_{f}) \cup {\cal P}_{e}$.  We now show that the
eigenvalues associated with $c$ and $c'$ are equal.  Indeed, because
of equation~(\ref{eq:pz}), the contribution of ${\cal P}_{f}$ in
$A_c(Y)$ and the contribution of ${\cal P}_{e}$ in $A_{c'}(Y)$ are
both equal to 1 and therefore $A_c(Y) = A_{c'}(Y)$, 
the other packages contained in $c$ and $c'$  being  the same.   Thus $Y_c$ is
also a solution of the Bethe equation associated with the set $c'$,
i.e., $A_{c'}(Y_c) = Y_c$.
Besides, thanks to equation~(\ref{eq:sz}), we notice that the
contribution of ${\cal P}_{f}$ to the eigenvalue $E_c$ is equal to the
contribution of ${\cal P}_{e}$ to $E_{c'}$.  Thus,  because
 the other packages contained in $c$ and $c'$  are the same, 
 we conclude from
equation~(\ref{eq:defEc}) that $E_c = E_{c'}$.  However, the
corresponding eigenvectors are different:  some of the chosen $z_i$'s being
different for $c$ and $c'$, the functions $\psi_c(x_1,\ldots, x_N)$
and $\psi_{c'}(x_1,\ldots, x_N)$ are not equal for the $\Omega$
different configurations. We have thus obtained a degenerate
eigenvalue $E_c = E_{c'}$  associated with   two different sets $c$
and $c'$.

To summarise, an eigenvalue corresponding to a choice set $c$ is
degenerate if there exists at least one package ${\cal P}_{f}$
entirely contained in $c$ and at least one package ${\cal P}_{e}$ that
does not intersect with $c$.  The fundamental reason is that a full
package of $Z_i$'s does not contribute to the Bethe equations and adds
up to a constant contribution in the eigenvalue formula.  Therefore
exchanging the packages ${\cal P}_{f}$ and ${\cal P}_{e}$ leads to the
same eigenvalue but not to the same eigenvector and results in
degeneracies in the spectrum.

As the size of the packages is $l$, we note that degeneracies are
possible only if $l \le N \le L-l$ with $l = L/\gcd(L,N)$.

\section{Combinatorial formulae for the degeneracies}

\label{s:oto}
We shall now enumerate  the degeneracies in the spectrum of the Markov
matrix $M$ of the TASEP with $N$ particles evolving on a ring of $L$
sites.  We assume the following {\em one-to-one hypothesis}: for each
choice set $c$ (among the $\Omega$ possible sets), the
self-consistency equation~(\ref{eq:acy=y}) has a unique solution $Y_c$
that provides the eigenvalue $E_c$ and the eigenvector $\psi_c$.  We
further assume that these eigenvectors are linearly independent; this
hypothesis, combined with the fact that the dimension of the
configuration space is $\Omega$, implies that the Bethe equations
provide a complete basis of eigenvectors.  We emphasize that the
one-to-one hypothesis is stronger than the assumption that the Bethe
basis is complete.  We have observed numerically on small size systems
that the functions $A_c(Y)$ are usually contraction mappings (which
would imply the one-to-one hypothesis), but we have not succeeded yet
to obtain a rigorous proof.  However we will see that this one-to-one
hypothesis allows to count the degeneracies and that the results are
in perfect agreement with numerical diagonalisations.  Thus we are
convinced that this hypothesis, or a weaker hypothesis with the same
consequences, is true and that it should be possible to prove it.

With this one-to-one hypothesis, counting degeneracies becomes merely
an exercise in combinatorics.  We first introduce some notations.  We
recall (see Section~\ref{s:package}) that $c = \{c_1, \dots, c_N \}$
is a set of $N$ integers chosen amongst $\{1, \dots, L \}$; moreover
$\{1, \dots, L \}$ is partitioned in $p$ packages ${\cal P}_1, \dots,
{\cal P}_p$, each containing $l = L/p$ integers with $p = \gcd(L,N)$ 
(see equations(\ref{eq:defp}--\ref{eq:defpack})).
 For a given set $c$ and for $0 \le i \le l$, we denote by $a_i$ the
number of packages ${\cal P}_k$ with $i$ elements in $c$ (i.e., such that
 ${\cal P}_k \cap c $  has $i$ elements).  Thus $a_0$ is
the number of  packages that do not  intersect $c$:  
 such packages will be referred to as   `empty' packages. 
 The number of the `full' packages  (i.e.,
entirely included in $c$)  is $a_l$.
 We call {\em partial} packages those that
are neither empty nor full.  Following this definition, we have
\begin{equation}
   a_i \ge 0
   \ \ , \ \ 
   \sum_{i=0}^l a_i = p 
   \ \ , \ \ 
   \sum_{i=0}^l i \, a_i = N  \, ,
   \label{eq:ai}
\end{equation}
$p$ being the total number of packages and $N$ the cardinality  of the set
$c$.  We call $a = (a_0, a_1, \dots, a_l)$ an {\em admissible
  partition} if it satisfies the  relations in
equation~(\ref{eq:ai}).  Equivalently an admissible partition
corresponds to a partition of the integer $N$ in which each term is
$\le l$ and which contains at most $p$ terms.

  The total number $\omega(a)$ of choice sets $c$ corresponding 
 to  a given admissible partition $a$  is 
\begin{equation}
   \omega(a) =  
   \frac{p!}{a_0!a_1! a_2!\dots a_l!} {l \choose 0}^{\displaystyle a_0}
          {l \choose 1}^{\displaystyle a_1} {l \choose 2}^{\displaystyle a_2} 
      \dots {l \choose l}^{\displaystyle a_{l}}    =  
       p!  \prod_{i=0}^l \frac{1}{a_i!} \ 
       {l \choose i}^{\displaystyle a_i}  \, .
  \label{eq:atoc}
\end{equation}
In this equation, the first factor enumerates the number of choices
for each type of packages among the $p$ available packages. The
factors of the type $ {l \choose i}^{a_i}$ give the number of choices
of $i$ elements among $l$ for each of the $a_i$ packages. According to
the one-to-one hypothesis, $\omega(a)$ represents also the number of
eigenvalues associated with the admissible partition $a$.

We have shown at the end of the previous section that two sets $c$ and
$c'$ provide a degenerate eigenvalue $E_c = E_{c'}$ if they are built
from the same partial packages and differ only by
 the selected  empty and full packages.  Thus  the eigenvalue
$E_c$  is $d(a)$ times degenerate with
\begin{equation}
   d(a) = {a_0 + a_l \choose a_l}  \, .  
   \label{eq:da}
\end{equation}
This relation is obtained by enumerating all the choice sets $c'$
obtained from $c$ by keeping the partial packages unchanged and
choosing $a_l$ full packages from the remaining  $a_l+a_0$ 
 packages.  We remark that $c$ and $c'$ correspond to the same
admissible partition $a$.  We also emphasize that the degeneracy
order depends only on 
 $a$ and not on the precise choice set   $c$. To resume, 
 a single order of  degeneracy  $d(a)$ is associated with 
the admissible partition $a$. 

Consequently, the $\omega(a)$ eigenvalues corresponding to  the
admissible partition $a$ form $m(a)$ multiplets of $d(a)$ degenerate
eigenvalues, where $m(a)$ is given by
\begin{eqnarray}
   m(a) =   \frac{ \omega(a) } { d(a) }
       =   \frac{p!}{a_1! a_2!\dots a_{l-1}! (a_0+a_l)!} 
          {l \choose 1}^{\displaystyle a_1} {l \choose 2}^{\displaystyle a_2} 
         \dots {l \choose l-1}^{\displaystyle a_{l-1}} \, .
       \label{eq:ma}
\end{eqnarray}
Because the value of $E_c$ depends only on the roots $Z_i$ belonging
to the partial packages, we remark that this equation can also be
obtained by enumerating the number of choices for these roots: the
first factor counts the number of choices for the partial packages
among the $p$ packages and the other factors enumerate the choices of
$i$ elements among $l$ for each of the $a_i$ partial  packages.

In order to know the total number $m(d)$ of multiplets with degeneracy
of order $d$, we must sum over all admissible partitions $a$ with
$d(a) = d$,
\begin{equation}
   m(d) = \sum_{a;\, d(a) = d} m(a).
   \label{eq:md}
\end{equation}

In the particular case of {\em half-filling} ($L=2N$), we have $p=N$
and $l=2$, $n=1$. The admissible partitions and the corresponding
degeneracies are parameterised by $a_0$: equation~(\ref{eq:ai}) leads
to $a_2=a_0$, $a_1=N-2a_0$ and $d(a) = {2a_0 \choose a_0}$.  The
relation between the admissible partitions and orders of 
degeneracies is therefore one-to-one
 (i.e., two different admissible partitions have different orders of 
 degeneracies): thus   the sum in
equation~(\ref{eq:md}) reduces to a single term.  Explicit formulae
and numerical results are given in (Golinelli and Mallick 2004).

  For fillings other than 1/2, we have $l > 2$
and two different admissible partitions can lead to the same order of
degeneracy.  Equation~(\ref{eq:md}) can not be further simplified 
 and the enumeration of admissible
partitions seems mandatory.  Nevertheless, we can verify the sum rule
\begin{equation}
   \Omega =  \sum_{d \ge 1} d \ m(d) = 
    \sum_{a} d(a) \ m(a) =  \sum_{a}  p! \prod_{i=0}^l \frac{1}{a_i!} \ 
       {l \choose i}^{\displaystyle a_i} 
    \label{eq:sr}
\end{equation}
where the last sum runs over the admissible partitions and $\Omega$,
  defined in equation~(\ref{eq:defomega}),  is
the size of the Markov matrix.   We use the identity
\begin{equation}
    (x+1)^L = \left[(x+1)^l\right]^p  
    = \left[ \sum_{i=0}^l {l \choose i} \ x^i \right]^p
    = \sum_{a_0, \dots, a_l} p! \prod_{i=0}^l \frac{1}{a_i!} \ {l \choose
    i}^{a_i} x^{ia_i}
\end{equation}
where $\sum_i a_i = p$.  We remark that $\Omega$ is the coefficient of
$x^N$ in $(x+1)^L$, whereas the coefficient of $x^N$ on the r.h.s.  is
precisely the number of admissible partitions  defined in
equation~(\ref{eq:ai})  and is identical to the r.h.s of
equation~(\ref{eq:sr}).

\begin{table}  \centering
  \begin{tabular}{c|rrrr| r r}
   packages & $a_0$& $a_1$ & $a_2$& $a_3$ & $d(a)$ & $m(a)$ \\
   \hline
  \includegraphics[width=\packagewidth]{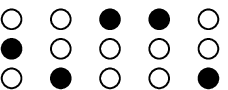} 
   & 0 & 5 & 0 & 0 &   1 &  243  \\
  \includegraphics[width=\packagewidth]{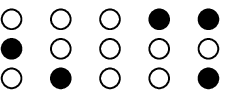} 
   & 1 & 3 & 1 & 0 &   1 &  1620 \\
  \includegraphics[width=\packagewidth]{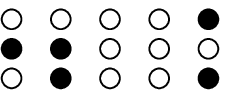} 
   & 2 & 1 & 2 & 0 &   1 &   810 \\
    \hline
  \includegraphics[width=\packagewidth]{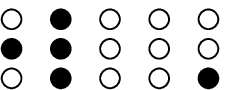} 
  & 2 & 2 & 0 & 1 &   3 &   90 \\
  \includegraphics[width=\packagewidth]{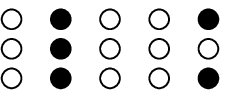} 
  & 3 & 0 & 1 & 1 &   4 &   15 \\
\end{tabular}
  \caption{
    An example of calculation of the degeneracies with $L=15$ and $N=5$.
     Each  row 
    describes an admissible partition $a=(a_0,\dots,a_l)$
    where $a_i$ counts the packages with $i$ chosen roots.
    In the first  column of this table, we have drawn one example
    of a choice set corresponding to $a$ that selects $N =5$ roots
     amongst  the $p=5$ packages of $l=3$  roots each.  
     The  eigenvalues corresponding to $a$ form $m(a)$
    multiplets of  order of degeneracy  $d(a)$, where $d(a)$ and $m(a)$
    are given by equations~(\ref{eq:da},
    \ref{eq:ma}).  The number of singlets ($d=1$) is obtained by summing
    the first three contributions, so 2673 singlets.
  }
  \label{tab:15-5}
\end{table}

In Table \ref{tab:15-5}, the explicit example $L=15$ and $N=5$ is worked
out.  We list the admissible partitions, calculate the corresponding order
of degeneracy from formula~(\ref{eq:da}) and enumerate the corresponding
multiplets by using equation~(\ref{eq:ma}).

These results are invariant under  `particle - hole' exchange: the
exclusion process with $N$ particles jumping to the right can be
mapped to a system with $L-N$ particles jumping to the left after
performing a particle - hole exchange.  Of course the spectrum of the
Markov matrix does not depend on the jumping direction. Thus, we know
a priori that the Markov matrices of the TASEP with $N$ and $L-N$
particles have  the same spectrum. We now verify this symmetry on the
formulae derived above.  Denoting with a `tilde' the quantities for
the model with $\tilde{N} = L-N$ particles, we find that $p =
\tilde{p}$ (because $\gcd(L,N) = \gcd(L-N,N)$), $\tilde{l}=l$ and
$\tilde{n} = l-n$.  By a particle - hole exchange, a partition $a =
(a_0, \dots, a_l)$ is transformed to $\tilde{a}$ where $\tilde{a_i} =
a_{l-i}$.  Hence, according to equations~(\ref{eq:da}, \ref{eq:ma}),
we obtain $d(\tilde{a}) = d(a)$ and $m(\tilde{a}) = m(a)$, i.e.,
degeneracies are indeed invariant by the particle - hole exchange.

\section{Discussion}

In this section, we draw some consequences of
equations~(\ref{eq:ai}-\ref{eq:md}).

We first verify that these equations always predict the existence of
singlets (i.e., isolated eigenvalues with degeneracy $d=1$).  In
equation~(\ref{eq:da}), we see that $d=1$ if and only if the partition
has no empty package ($a_0=0$) or no full package ($a_l=0$) or if both
$a_0 = a_l =0$.  For instance, the stationary
eigenvalue 0 is always a singlet thanks to Perron-Frobenius theorem:
in fact, the choice set for the stationary  state is $c = \{1,\dots, N\}$
and hence each package ${\cal P}_k$ has $n = N/p$ selected elements.
The corresponding partition is thus given by $a_i = 0$ for $i \ne n$
and $a_n = p$: this implies $d=1$, as expected.

\begin{table}  \centering
  \begin{tabular}{rr|rrrrr}
   $L$   & $N$ &  $m(1)$ &  $m(2)$ & $m(6)$ & $m(20)$ & $m(70)$
\\ \hline
    2    &     1     &     2 \\
    4    &     2     &     4  &     1  \\
    6    &     3    &     8  &     6  \\
    8    &     4    &    16  &    24  &    1 \\
   10    &     5    &    32  &    80  &   10 \\
   12    &     6    &    64  &   240  &   60  & 1      &     \\
   14    &     7    &   128  &   672  &  280  & 14     &     \\
   16    &     8    &   256  &  1792  & 1120  & 112    &  1  \\
   18    &     9    &   512  &  4608  & 4032  & 672    & 18  \\
  \end{tabular}
  \caption{\em 
    Spectral degeneracies in the TASEP at filling $\rho = 1/2$: $L$ is the
    size of the lattice, $N$ the number of particles; the other columns
    give $m(d)$ the number of  multiplets with degeneracy $d$.
   }
  \label{tab:1/2}
\end{table}

\begin{table}  \centering
  \begin{tabular}{r|r r|rrrrrr}
   $\rho$ & $L$ & $N$ & $m(1)$ & $m(2)$ & $m(3)$ & $m(4)$ & $m(5)$ & $m(15)$
\\ \hline
   1/3    &  9  &  3  &     81 &        &     1   \\
          &  12 &  4  &    459 &        &    12   \\
          &  15 &  5  &   2673 &        &    90  &     15   \\
          &  18 &  6  &  15849 &        &   540  &    270 &        &   1 \\
          &  21 &  7  &  95175 &        &  2835  &   2835 &    189 &  21 \\
   \hline 
   1/4    &  16 &  4  &   1816 &        &        &      1 \\
          &  20 &  5  &  15424 &        &        &     20 \\
          &  24 &  6  & 133456 &        &        &    240 &     36  \\
   \hline 
   1/5    &  25 &  5  &  53125 &        &        &        &      1  \\
   2/5    &  15 &  6  &   4975 &     15 \\
  \end{tabular}
  \caption{\em 
    Examples of spectral degeneracies in the TASEP at filling $\rho \ne
    1/2$: $L$ is the size of the lattice, $N$ the number of particles; the
    other columns give $m(d)$ the number of multiplets with degeneracy $d$.
   }
  \label{tab:other}
\end{table}

In order to obtain degenerate eigenvalues i.e., admissible partitions
with $d \ge 2$, we must have $a_0 \ne 0$ and $a_l \ne 0$, i.e., at
least one empty package and one full package.  According to
equation~(\ref{eq:ai}), the existence of degeneracies is given by the
condition $l \le N \le L-l$ or equivalently by
\begin{equation}
   L \le pN \le pL-L
   \label{eq:lnl}
\end{equation}
where $p = \gcd(L,N)$.  Some numerical examples are given in Tables 
\ref{tab:1/2} and \ref{tab:other}.

We now  analyse the TASEP at a fixed value of the filling $\rho =
n/l$, $n$ and $l$ being mutually prime. The two integers $(L,N)$ are
parameterised by the single number $p$ defined in equation~(\ref{eq:defp}).
According to the condition~(\ref{eq:lnl}), degeneracies appear when
\begin{equation}
   p \ge \max\left( \frac{1}{\rho}, \frac{1}{1-\rho} \right) \, .
\end{equation}
This condition can  always be  fulfilled when $0 < \rho < 1$. 
  Moreover, when the
system size $L$ and the number of particles $N$ increase with a given
rational $\rho$, higher and higher orders of degeneracies appear and become
generic in the large $p$ (or $L$) limit.  The admissible partition $a=(a_0,
\dots, a_l)$ that maximises $ \omega(a) = d(a) \, m(a)$ (i.e., the total
number of eigenvalues associated with $a$) is given by
\begin{equation}
   a_i \sim p {l \choose i} \rho^i (1-\rho)^{l-i}
   \ .
 \label{opt1}
\end{equation}
The corresponding order of degeneracy increases exponentially with the
size $L$
as $d \propto D^L$ where
\begin{equation}
    D =  \left(  1 +\frac{v}{u} \right)^{u/l} 
       \left(1 +\frac{u}{v}  \right)^{v/l} 
   \label{eq:d}
   \ \mbox{with} \ 
   u = \rho^l 
   , \ 
   v = (1-\rho)^l
   .
\end{equation}
 For example, $D_{1/2}=2^{1/4} \approx 1.189$ for $\rho = 1/2$;
$D_{1/3} = (9^3/2^8)^{1/27} \approx 1.040$ for $\rho = 1/3$, etc...
$D$ converges rapidly to 1 when the denominator $l$ of $\rho$ grows.
That explains why, in numerical studies of systems of limited size,
degeneracies are found only when $l$ is rather small, i.e., when
$\rho$ is a `simple' fraction.

Similarly, we can also determine the admissible partition $a$ that
maximises $m(a)$, the number of multiplets. This optimal partition has
$a_l=0$ when $\rho < 1/2$, and $a_0=0$ when $\rho > 1/2$. In either case,
this corresponds to $d=1$.  Thus the most numerous multiplets for $\rho \ne
1/2$ are singlets. This result does not contradict equations~(\ref{opt1},
\ref{eq:d}) in which the product $d(a)\, m(a)$ is maximised.  For the
special case, $\rho = 1/2$, the partition that maximises $m(a)$ has both
$a_0$ and $a_l$ different from 0 and the corresponding order of degeneracy
increases as $d \propto 2^{L/6}$ (Golinelli and Mallick 2004).

  Furthermore, for a given number $N$ of particles, we can search the
values of $L$ where degeneracies appear.  Because of the particle -
hole  symmetry (i.e., $N \Leftrightarrow L-N$), we need to consider
only the case $N \le L/2$. Then, the condition~(\ref{eq:lnl}) becomes
$2N \le L \le pN$.  Because $p \le N$, this implies that $L \le N^2$:
only a finite number of $L$ values are possible. 
In the dilute limit (when $L$ becomes large and $N$ remains fixed),
the TASEP  has thus no degeneracy.

All these results have been derived on the basis of the `one-to-one'
hypothesis stated at the beginning of section~\ref{s:oto}. It is
therefore crucial to compare our formulae with numerical results.
We have numerically diagonalised the Markov matrix of the TASEP for
certain values of the parameters $(L,N)$.  We use the translation
symmetry to split the matrix of size $\Omega$ into $L$ matrices of
size about $\Omega/L$. The spectrum is then computed by using Lapack
library (Anderson et al 1999) and degeneracies are counted; details
about this procedure are given in (Golinelli and Mallick 2004).  We
have investigated systematically all the systems $(L,N)$ with $L \le
19$.  For $L \ge 20$, we have studied the systems $(L,N)$ in which
non-trivial degeneracies are predicted and in which the size of the
diagonalised matrix remains less than 6000.  Results are given in
Table~\ref{tab:1/2} for $\rho = 1/2$ and in Table~\ref{tab:other} for
other values of $\rho$.  The numerical results are in perfect
agreement with our analytical predictions, 
equations~(\ref{eq:da}-\ref{eq:md}).

The degeneracies for systems much larger than those listed in Tables
\ref{tab:1/2} and \ref{tab:other} can be calculated from the
formulae~(\ref{eq:da}-\ref{eq:md})  for systems with several
hundred sites and particles.  However, the full numerical
diagonalisation of the Markov matrix consumes a computer time of the
order of $\Omega^3 \propto (\rho^\rho (1-\rho)^{1-\rho})^{-3L}$.  Such
a fast growth limits the comparison between numerical diagonalisations
and the exact formulae.

\section{Conclusion}

The spectrum of the Markov matrix of the TASEP on a one-dimensional
periodic lattice has a rich structure with multiplets having
degeneracies of high order. This structure depends on the filling
fraction and presents arithmetical properties related to some
particular partitions of the total number of particles. We have
derived analytical formulae for the spectral degeneracies by analysing
the Bethe equations for the TASEP.  These predictions have been
verified by numerical calculations  and we conjecture that the formulae
we propose are exact. Our arguments are based on a `one-to-one
hypothesis' which is stronger than assuming the completeness of the
Bethe Ansatz. Although the agreement between numerical results and
analytical predictions is a strong argument in favour of the
one-to-one hypothesis, it is possible that this hypothesis is not
satisfied but that a weaker formulation, leading to the same spectral
structure, holds good.  We plan to study the completeness of the Bethe
Ansatz and this one-to-one hypothesis more precisely in a future work.

Our derivation of the spectral structure from the Bethe equations is
rather indirect. In fact, the presence of such high orders of
degeneracies is a strong evidence for hidden symmetries in the model.
In other words, the TASEP should be invariant under a group operating
on the configuration space such that each multiplet of a given order
of degeneracy is an irreducible representation of this group. The
orders of degeneracies would then classify the dimensions of
irreducible representations and the number of multiplets of a given
order would represent the multiplicity of a given irreducible
representation in the global representation over the configuration
space. By using the techniques of the algebraic Bethe Ansatz, we have
constructed a series of nonlocal operators that act on multiple
particles and that commute with the Markov matrix. We hope that a
study of the representations of these operators will allow us to
understand the structure of the spectrum in a purely algebraic manner
without having to analyse the solutions of the Bethe equations.

The TASEP evolution can be generalized by introducing  a fugacity parameter
$\lambda$, i.e., by multiplying each non-diagonal term of the Markov
matrix by a factor $\lambda.$ This parameter has been used to
calculate the large deviation function of the total
 particle displacement  (Derrida and
Lebowitz 1999). We verified that the spectral degeneracies for the
`TASEP + fugacity' model are the same as those found for TASEP. This
is not surprising because the properties of the Bethe equations are
not altered by introducing a fugacity and the arguments given in
sections 4 and 5 can be generalized without any difficulty. However,
in the case of the {\it partially}  asymmetric exclusion process, in which
particles can jump in both directions, the spectrum has a much simpler
structure: eigenvalues are either singlets or doublets. Lastly, for
the TASEP on an open lattice, the spectrum is made only of singlets.

\subsection*{Acknowledgements}
  
 We thank   M. Gaudin and V. Pasquier for useful discussions. We are
 thankful to S. Mallick for a critical reading of the manuscript. 

\appendix 
\section*{Appendix:  Layout of  the solutions of the Bethe equations}

In this Appendix, we explain how to label the $L$ solutions
$(Z_1,\dots, Z_L)$ of the polynomial equation of degree $L$
\begin{equation}
   (1-Z)^N (1+Z)^{L-N} = Y \, ,
  \label{eq:zzy}
\end{equation}
($0 \le N \le L$), in such a way that each $Z_k(Y)$ is an analytic
function of the parameter $Y$ in the complex plane with a branch cut
along the real semi-axis $[0,+\infty)$.

\begin{figure}
  \centering \includegraphics[width=\figwidth, keepaspectratio] {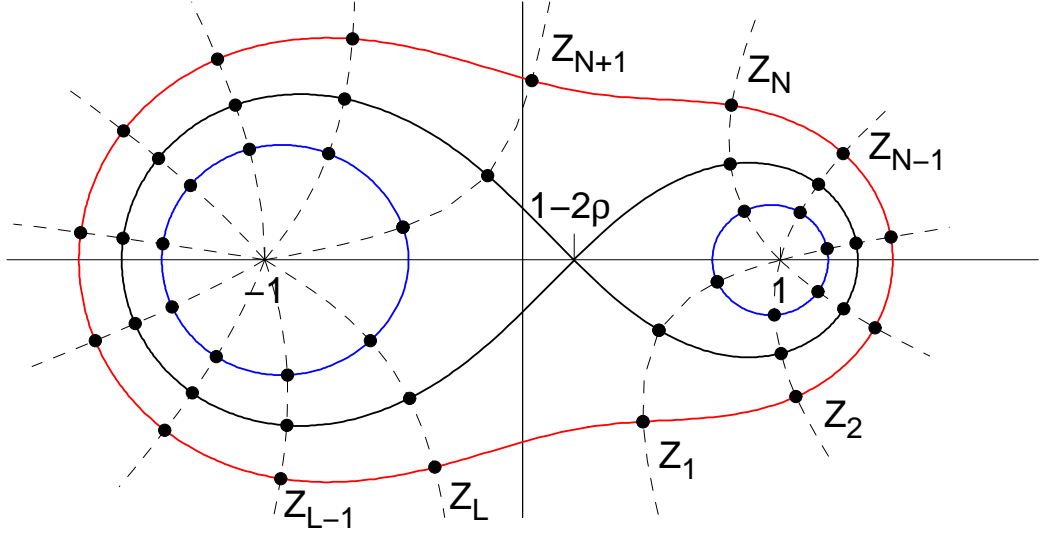}
  \caption{\em 
    Labelling the roots of the Bethe equations.  Here $L = 15$, $N=
    6$, $\phi = \pi/2$ and $r/r_c = 0.8,1, 1.2$ (see text).  The
    (resp. blue, black and red) 
    continuous curves are the
    corresponding Cassini ovals.  When $r$ is fixed and $\phi$ varies
    from 0 to $2\pi$, each $Z_k$ slips counterclockwise along a part
    of the Cassini oval.  When $\phi$ is fixed and $r$ varies from
    $\infty$ to 0, each $Z_k$ travels along a dashed curve from
    $\infty$ to points +1 or -1.  }
  \label{fig:cassini}
\end{figure}

A non-zero complex number $Y$ can be written in a unique way as
\begin{equation}
   Y = r^L \, e^{i \phi} 
   \ \ \mbox{with}\ \  
   0 \le \phi < 2 \pi
   \ ,
\end{equation}
$r$ being a positive real number.  This determination of the argument
has a branch cut along $[0,+\infty)$.  For a given value of $r$, the
complex numbers $Z_k$ belong to the generalized Cassini oval defined
by
\begin{equation}
   |Z-1|^{\rho} |Z+1|^{1-\rho} = r
\end{equation}
where $\rho = N/L$ is the filling of the system.  As shown in
 Fig.~\ref{fig:cassini}, the topology of the Cassini oval depends on
the value of $r$ with a critical value
\begin{equation}
   r_c = 2 \rho^\rho(1-\rho)^{1-\rho}
   :
\end{equation}
\begin{itemize}
\item for $r < r_c$, the curve consists of two disjoint ovals with $N$
  solutions on the oval surrounding $+1$ and $L-N$ solutions on the
  oval surrounding $-1$.
  
\item for $r = r_c$, the curve is a deformed Bernoulli lemniscate
  with a double point at $Z_c = 1 - 2 \rho$.
  
\item for $r > r_c$, the curve is a single loop with $L$ solutions.
\end{itemize}
The Cassini ovals are symmetrical only if $\rho = 1/2$.

In order to label the solutions, we start by considering the limit $r
\to \infty$ for a given $\phi$.  Equation~(\ref{eq:zzy}) then becomes
\begin{equation}
  Z^L \sim r^L \, \exp [i(\phi - N\pi)]  \, .  
\end{equation}
The solutions $Z_k$ are labelled by
\begin{equation}
   Z_k \sim r \, \exp\left[ \frac{i}{L} [\phi - N\pi + 2(k-1)\pi] \right]
   \ \ \mbox{with}\ \  
    k = 1, \ldots,  L
   \, .
   \label{eq:zk}
\end{equation}
In other words, the $Z_k$ are regularly distributed along a large
circle of radius $r$ with
\begin{equation}
   \frac{\phi - N\pi}{L} \le \arg Z_1 < \dots < \arg Z_L < 
    \frac{\phi - N\pi}{L} + 2\pi
    \, .
\end{equation}

This labelling, obtained for large $r$, is extended by analytic
continuation to all values of $r$, keeping $\phi$ {\em fixed}.  The
loci of the $Z_k$ are drawn in Fig.~\ref{fig:cassini} (dashed curves):
they are orthogonal to the Cassini ovals.  A singularity appears along
the branch $\phi = 0$ because $Z_1$ and $Z_{N+1}$ collapse into each
other at the double point $Z_c$ when $r=r_c$; we circumvent it by
choosing $\phi = 0^+$.

With this labelling, the solutions are ordered along the Cassini
ovals.  Moreover, when $r < r_c$, the solutions $(Z_1,\dots,Z_N)$
group together on the right oval and $(Z_{N+1}, \dots, Z_L)$ on the
left oval.

\section*{References}

\begin{itemize}
  
\item Anderson E. and al., 1999, {\em LAPACK Users' Guide},
  (Philadelphia, SIAM)

\item Derrida B., 1998, {\em An exactly soluble non-equilibrium
    system: the asymmetric simple exclusion process}, Phys. Rep.  {\bf
    301}, 65.

\item Derrida B., Evans M.~R., Hakim V., Pasquier V., 1993, {\em Exact
    solution of a 1D asymmetric exclusion model using a matrix
    formulation}, J. Phys. A: Math. Gen. {\bf 26}, 1493.

\item Derrida B.  and Lebowitz J.~L., 1998, {\em Exact large deviation
    function in the asymmetric exclusion process}, Phys. Rev. Lett.
  {\bf 80}, 209.

\item Derrida B., Lebowitz J.~L., Speer E.~R., 2003, {\em Exact large
    deviation functional of a stationary open driven diffusive system:
    the asymmetric exclusion process}, J. Stat. Phys. {\bf 110}, 775.

\item Dhar D., 1987, {\em An exactly solved model for interfacial
    growth}, Phase Transitions {\bf 9}, 51.

\item Evans M.~R., Kafri Y., Koduvely H.~M., Mukamel D., 1998, {\em
    Phase separation in one-dimensional driven diffusive systems},
  Phys. Rev. Lett.  {\bf 80}, 425.

\item Gaudin M., 1983, 
 {\em  La Fonction d'Onde de Bethe}, (Masson, Paris).
  
\item Gaudin M. and Pasquier V., 2004, {\em private communication}.

\item Golinelli O. and Mallick K., 2004, {\em Hidden symmetries in the
    asymmetric exclusion process}, JSTAT P12001; cond-mat/0412353.

\item Gwa L.-H., Spohn H., 1992, {\em Bethe solution for the
    dynamical-scaling exponent of the noisy Burgers equation}, Phys.
  Rev. A {\bf 46}, 844.

\item Kim D., 1995, {\em Bethe Ansatz solution for crossover scaling
    functions of the asymmetric XXZ chain and the
    Kardar-Parisi-Zhang-type growth model}, Phys. Rev. E {\bf 52},
  3512.

\item Sch\"utz G.M., 2001 Phase Transitions and Critical Phenomena,
  Vol. 19, (Academic, London).

\end{itemize}

\end{document}